\pdfoutput=1
\documentclass{article}

\PassOptionsToPackage{numbers,sort&compress}{natbib}
\usepackage[final,dblblindworkshop]{neurips_2026}
\workshoptitle{Resource-Aware Agentic AI}

\usepackage[utf8]{inputenc}
\usepackage[T1]{fontenc}
\usepackage{amsmath}
\usepackage{amssymb}
\usepackage{array}
\usepackage{booktabs}
\usepackage{graphicx}
\usepackage{hyperref}
\usepackage{microtype}
\usepackage{multirow}
\usepackage{nicefrac}
\usepackage{xcolor}
\usepackage{xspace}
\usepackage{url}
\hypersetup{
  hidelinks,
  pdftitle={The Physical Cost of Context: Quality-Constrained Resource
    Accounting for Coding Agents},
  pdfauthor={Zhongming Yu, Yitong Ma, Hengjia Yu, Boqin Yuan, Xiangye Lin,
    Hejia Zhang, Jishen Zhao},
}

\definecolor{better}{HTML}{009E73}
\definecolor{worse}{HTML}{D55E00}
\definecolor{RuleGray}{HTML}{555555}
\newcommand{\savingsign}[1]{\ifx-#1\textminus\else#1\fi}
\newcommand{\saving}[1]{{\footnotesize\textcolor{better}{(\savingsign#1)}}}
\newcommand{\increase}[1]{{\footnotesize\textcolor{worse}{(#1)}}}
\newcommand{\vsbase}[1]{{\footnotesize\textcolor{RuleGray}{#1}}}

\newcommand{\PoweredGrepRecall}{.470}

\newcommand{\PoweredEagerRecall}{.478}

\newcommand{\PoweredCompactRecall}{.494}

\newcommand{\PoweredRetrieverRecallFive}{.431}
\newcommand{\PoweredRetrieverRecallTen}{.548}
\newcommand{\PoweredEagerDiffPoints}{+0.8}
\newcommand{\PoweredEagerCILowPoints}{-2.6}

\newcommand{\PoweredCompactDiffPoints}{+2.4}
\newcommand{\PoweredCompactCILowPoints}{-0.9}

\newcommand{\PoweredCompactGate}{meets}

\newcommand{\CacheOnCompactUncachedChange}{-38.2}

\newcommand{\CacheEffectEagerPromptChange}{+8.3}
\newcommand{\CacheEffectEagerPromptCILow}{-5.7}
\newcommand{\CacheEffectEagerPromptCIHigh}{+24.7}

\newcommand{\CacheEffectEagerUncachedChange}{-67.7}
\newcommand{\CacheEffectEagerUncachedCILow}{-71.0}
\newcommand{\CacheEffectEagerUncachedCIHigh}{-63.8}

\newcommand{\CacheCompactVsEagerUncachedKInteraction}{-7.1}
\newcommand{\CacheCompactVsEagerUncachedKInteractionCILow}{-17.3}
\newcommand{\CacheCompactVsEagerUncachedKInteractionCIHigh}{+2.3}
\newcommand{\CacheCompactVsEagerWallSecondsInteraction}{+0.1}
\newcommand{\CacheCompactVsEagerWallSecondsInteractionCILow}{-5.4}
\newcommand{\CacheCompactVsEagerWallSecondsInteractionCIHigh}{+5.7}

\newcommand{\MiniRepositoryClusters}{12}

\newcommand{\MiniEagerResolved}{29}
\newcommand{\MiniEagerAttempts}{100}

\newcommand{\MiniMaskResolved}{26}
\newcommand{\MiniMaskAttempts}{100}

\newcommand{\MiniMaskDiffPoints}{-3.0}
\newcommand{\MiniMaskCILowPoints}{-8.6}

\newcommand{\MiniMaskQualityGate}{does not meet}

\newcommand{\MiniMaskPromptChange}{-33.3}

\newcommand{\MiniMaskCallsChange}{+2.6}

\newcommand{\MiniMaskUncachedChange}{+608.5}

\newcommand{\MiniMaskWallChange}{+36.9}

\newcommand{\MiniMaskEnergyChange}{+125.6}

\newcommand{\MiniMaskFirstCallMedian}{6}

\newcommand{\MiniValidatedPreloadCells}{300}
\newcommand{\MiniValidatedNoSeedCells}{100}
\newcommand{\MiniRecoveryInvalidatedCells}{13}
\newcommand{\MiniRecoveryDiscardedRequests}{277}
\newcommand{\MiniRecoveryDiscardedMinutes}{20.8}
\newcommand{\CacheInstrumentedBlocks}{150}
\newcommand{\CacheInstrumentedHours}{7.90}
\newcommand{\CacheInstrumentedEnergyKWh}{0.183}
\newcommand{\MiniInstrumentedBlocks}{32}
\newcommand{\MiniInstrumentedHours}{14.65}
\newcommand{\MiniInstrumentedEnergyKWh}{0.281}

\newcommand{\AddedInstrumentedHours}{22.55}
\newcommand{\AddedInstrumentedEnergyKWh}{0.464}

\newcommand{\evidencefigurepath}{figures/evidence_dashboard.pdf}
\title{The Physical Cost of Context: Quality-Constrained Resource Accounting
for Coding Agents}

\author{%
  Zhongming Yu\textsuperscript{1}\quad
  Yitong Ma\textsuperscript{1}\quad
  Hengjia Yu\textsuperscript{1}\quad
  Boqin Yuan\textsuperscript{1}\\
  \bfseries
  Xiangye Lin\textsuperscript{2}\quad
  Hejia Zhang\textsuperscript{1}\quad
  Jishen Zhao\textsuperscript{1}\\[3pt]
  \normalfont\small
  \textsuperscript{1}University of California, San Diego\qquad
  \textsuperscript{2}University of Massachusetts Amherst%
}

\begin{document}

\maketitle

\begin{abstract}
Coding agents resend a growing context on every model call, and context
policies are usually judged by the prompt tokens they save.  Executed work also
depends on call count and prefix reuse: a caching runtime prefills only the
tokens after the longest cached prefix.  A shorter
prompt can therefore cost more if it rewrites cached history.  Our
quality-gated protocol splits executed prefill into logical prompt minus reused
prefix and measures KV residency, time, and energy.  On 450 localization cases
derived from SWE-rebench V2, preloading ten retrieved blocks (\textsc{Eager})
and adding anchored history compaction (\textsc{Compact}) both pass a
five-point non-inferiority gate.  In a separate 900-cell study, \textsc{Eager}
cuts model calls by 35.7\% and the GPU-energy proxy by 33.2\% against
on-demand search (\textsc{Grep}); \textsc{Compact} saves nothing further.
Enabling prefix caching leaves \textsc{Eager}'s prompt volume similar
($+8.3\%$; its 95\% CI spans zero) but cuts its uncached prefill by 67.7\%.
Deleting history can turn a token saving into extra work.  With mini-SWE-agent
on 100 SWE-bench Verified instances, keeping only the last five observations
(\textsc{Mask-5}) cuts \textsc{Eager}'s prompt volume by 33.3\% but loses
135.9k reusable tokens per task.
Uncached prefill rises by 608.5\%, the energy proxy by 125.6\%, and the policy
misses the quality gate.  Policies thus save physical work by shortening
the trajectory and can multiply it by editing reusable history.
\end{abstract}

\section{Introduction}
\label{sec:introduction}

Repository localization maps an issue report to files, classes, and functions.
Retrieval methods rank files using report--code similarity, structure, and
history~\citep{buglocator,bluir,learningtorankbug}.  Modern agents combine
iterative retrieval~\citep{repocoder}, structure-aware search
\citep{autocoderover,agentless,orcaloca}, or graph navigation
\citep{locagent,cosil,repograph,codexgraph}.

These agents repeatedly search, read, and reassess the retrieved source.  A
ReAct-style loop~\citep{react} turns a short issue report into a trajectory of
tool outputs and intermediate hypotheses.  Larger contexts may improve
evidence coverage, but every retained observation can add input processing
before decoding (prefill) and occupy the model's key--value (KV) cache on later
turns.  On a local device, model weights, KV cache, retrieval services, and the
host process also share one memory and power budget.

Many context-management studies report tokens, task success, or provider
cost~\citep{llmlingua,longllmlingua,agentdiet,complexity-trap,swe-pruner}.
Token counts do not determine physical cost: equal reductions in logical tokens
can correspond to different amounts of executed prefill, depending on request
boundaries and prefix reuse~\citep{promptcache,cacheblend}.  On a
unified-memory device, model and host allocations also share one
pool~\citep{dgxspark}, so we track host allocation, paging, and latency
together.

Our agent returns answer blocks for an issue-like query.  A frozen
retrieval-augmented generation (RAG) path can insert ten retrieved code blocks
before the first turn.  We ask when logical-token savings predict physical savings,
and when cache or trajectory effects break that prediction.

\begin{figure}[t]
  \centering
  \includegraphics[width=\linewidth]{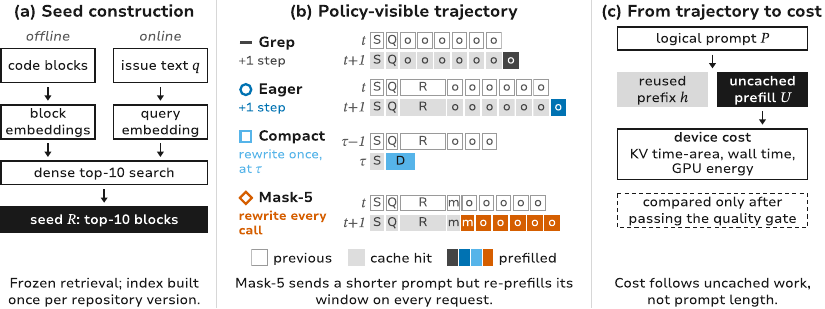}
  \caption{Experimental boundary from repository to device.  (a) Frozen dense
  retrieval selects top-10 blocks $R$.  (b) Policies alter evidence delivery and
  visible history.  (c) Request and prefix-reuse boundaries translate logical prompt $P$ into
  uncached work $U$ and device cost, which we compare only after checking quality.}
  \label{fig:supply-chain}
\end{figure}

Measuring this path (Figure~\ref{fig:supply-chain}) raises three challenges.
\textbf{C1: Trajectory attribution.}  Evidence supply, delivery time, and
history retention can each change what the agent does next.  Matched contrasts
hold the other mechanisms fixed, and we treat the resulting trajectory as an
outcome of the policy.  \textbf{C2: Logical-to-physical translation.}  A
policy changes both the number of requests and prefix reuse within each
request.
Logical tokens can therefore order policies differently from uncached prefill,
KV occupancy, latency, or energy.  \textbf{C3: Quality-constrained comparison.}
A resource comparison must first bound task-quality loss.

We study four fixed context policies on an NVIDIA DGX Spark: \textsc{Grep},
\textsc{Eager}, \textsc{Compact}, and \textsc{Mask-5}.  They differ in when
retrieved evidence arrives and which observations stay visible.  We make two
contributions:

\begin{enumerate}
  \item \textbf{A quality-gated physical-accounting protocol.}  The protocol
  combines paired task-quality bounds with model calls, uncached prefill, KV
  time-area, wall time, and idle-subtracted GPU energy.  It charges failures,
  reports index construction separately, and decomposes executed prefill into
  logical input minus reused prefixes.

  \item \textbf{A controlled policy analysis.}  Byte-identical candidates
  separate evidence delivery from history reduction.  On 450 real-issue
  localization cases derived from SWE-rebench V2, both preload policies
  satisfy the quality gate.  Instrumented
  runs show that eager evidence saves work by avoiding calls, anchored
  reduction adds no saving beyond \textsc{Eager}, and repeated observation
  masking can erase prompt savings, consistent with reduced prefix reuse.
\end{enumerate}

\section{Context as a Resource-Control Problem}
\label{sec:method}

\subsection{Two Paths from Context to Cost}

A context policy changes both the requests an agent makes and the prefixes its
runtime can reuse.  Serving runtimes such as vLLM and SGLang keep the KV
entries of earlier requests; when a new request begins with a cached token
prefix, the runtime reuses those entries and prefills only the remaining
suffix~\citep{vllm,sglang}.  Editing an early token therefore forces every later
token to be recomputed.  Let $n_p$ be the number of model requests under policy
$p$, $\ell_{p,t}$ the logical prompt tokens in request $t$, and $h_{p,t}$ the
tokens served from a reusable prefix.  Aggregate logical and executed input
work satisfy
\begin{equation}
  T_{\mathrm{prompt}}(p)=\sum_{t=1}^{n_p}\ell_{p,t},\qquad
  T_{\mathrm{uncached}}(p)=T_{\mathrm{prompt}}(p)-
  \underbrace{\sum_{t=1}^{n_p}h_{p,t}}_{T_{\mathrm{hit}}(p)}.
  \label{eq:cost-decomposition}
\end{equation}
For a treatment--control contrast, $\Delta T_{\mathrm{uncached}}=
\Delta T_{\mathrm{prompt}}-\Delta T_{\mathrm{hit}}$.  Prompt savings therefore
increase executed prefill when a policy loses more reusable tokens than it
removes.  Because $n_p$ also depends on the policy, fewer calls can
reduce total work even when each call is no cheaper.  We use this decomposition
to distinguish trajectory effects from runtime-reuse effects.

\subsection{Frozen Code-RAG Supply Path}

Following RAG~\citep{rag}, each snapshot is compiled offline into CodeNib
function/class blocks~\citep{codenib} and encoded by a revision-pinned
Qwen3-Embedding-0.6B model~\citep{qwenembedding}.  The same encoder maps the
issue query, and dense search returns ten spans.  We exclude lexical fusion,
graph expansion, and reranking.

Query encoding and search count toward each preload task; compilation is
amortized separately.  Hashes verify byte-identical preload inputs, and
\textsc{Grep} keeps the same tools without a seed
\mbox{(Figure~\ref{fig:supply-chain}a--b)}.

\subsection{Policy Algebra}

Let $\sigma\in\{\mathrm{loc},\mathrm{patch}\}$ identify the agent scaffold.
For scaffold $\sigma$, let $S_\sigma$ be its system prompt and
$H_{\sigma,t}=(a_1,o_1,\ldots,a_t,o_t)$ its canonical action--observation
history.  An experimental arm is $p=(\sigma,e,r)$, where
$e\in\{\varnothing,R_{10}\}$ is the evidence supplied before the first turn
and $r$ controls visible serialization.  With $M_E$ denoting the compiled
repository index and $R_{10}=R_{10}(q;M_E)$, the visible state and the
serialized model input are
\begin{equation}
  V_{\sigma,t}^{(e)}=q\oplus e\oplus H_{\sigma,t},\qquad
  P_{\sigma,t}^{(e,r)}=S_\sigma\oplus r\!\left(V_{\sigma,t}^{(e)}\right),
  \label{eq:policy-context}
\end{equation}
where $\oplus$ denotes serialization.  Policies retain the raw
$H_{\sigma,t}$ for audit and change only its model-visible serialization.

Three deterministic context-serialization operators cover the tested designs:
\begin{align}
  I\!\left(V_{\sigma,t}^{(e)}\right) &= V_{\sigma,t}^{(e)}, \nonumber\\
  A_\sigma\!\left(V_{\sigma,t}^{(R_{10})}\right) &=
    \begin{cases}
      V_{\sigma,t}^{(R_{10})}, & t<\tau,\\
      D_{\tau,\sigma}\oplus H_{\sigma,\tau+1:t}, & t\geq\tau,
    \end{cases} \nonumber\\
  L_5\!\left(V_{\mathrm{patch},t}^{(R_{10})}\right)
    &=q\oplus R_{10}\oplus
    H_{\mathrm{patch},t}\!\left[o_i\leftarrow\mu(|o_i|)
      \ \text{for }i\leq t-5\right].
  \label{eq:retention-operators}
\end{align}
Here $\tau$ is the first successful source read.  The scaffold-specific anchor
$D_{\tau,\sigma}$ contains the task query exactly once, the successful read,
and a bounded pre-rewrite assessment.  Following SWE-agent's
\texttt{LastNObservations}~\citep{sweagent}, $L_5$ keeps actions and tool IDs
but replaces older outputs with line-count markers.  Neither invokes a model.
$A_\sigma$ rewrites once; $L_5$ rewrites whenever an observation leaves the
window.  Both can change later actions and invalidate reuse, so total saving
depends on $n_p$ and $T_{\mathrm{hit}}(p)$ in
Equation~\eqref{eq:cost-decomposition}.

The arm definitions are therefore
\begin{equation}
  \begin{aligned}
    G&=(\mathrm{loc},\varnothing,I),
      & N&=(\mathrm{patch},\varnothing,I),\\
    E_\sigma&=(\sigma,R_{10},I),
      & C_\sigma&=(\sigma,R_{10},A_\sigma),
      & M_5&=(\mathrm{patch},R_{10},L_5).
  \end{aligned}
  \label{eq:policy-matrix}
\end{equation}
Thus $\mathcal{P}_{\mathrm{loc}}=\{G,E_{\mathrm{loc}},C_{\mathrm{loc}}\}$ and
$\mathcal{P}_{\mathrm{patch}}=\{N,E_{\mathrm{patch}},C_{\mathrm{patch}},M_5\}$.
Informally, \textsc{Grep} ($G$, localization) and native mini-SWE-agent
($N$, patching) start from the issue alone and keep their full history.
\textsc{Eager} ($E_\sigma$) prepends the ten retrieved spans and also keeps
the full history.  \textsc{Compact} ($C_\sigma$) matches \textsc{Eager} until
the first successful source read, then replaces the candidate list and
exploratory history with a short anchor.  \textsc{Mask-5} ($M_5$) keeps every
action but shortens each tool output older than the latest five to a
placeholder, so once the window fills, every request rewrites an observation
that the previous request had already sent.  Within a study, seeded arms receive byte-identical $R_{10}$.  The $C_\sigma-E_\sigma$ contrast
estimates the end-to-end effect of the anchor rewrite;
$M_5-E_{\mathrm{patch}}$ estimates masking.  Contrasts against $G$ or $N$ also
add the retrieved evidence, so they combine delivery with any retention change.
We do not pool absolute resources across Tables~\ref{tab:profile}
and~\ref{tab:mini-e2e}.

\subsection{Quality-Constrained Resource Objective}

For each gated treatment $p$, let $Q(p)$ denote its task-quality endpoint and
let $c(p)$ denote its registered control.  We use answer-block Recall@5 for
localization and official resolved rate for patch generation.  Let
\begin{equation}
  \mathbf{R}(p) = (T_{\mathrm{prompt}}, T_{\mathrm{output}}, N_{\mathrm{req}},
  T_{\mathrm{uncached}}, L_{\mathrm{TTFT}}, L_{\mathrm{wall}},
  K_{\int}, K_{\max}, U_{\max}, E_{\mathrm{gpu}})
  \label{eq:resource-vector}
\end{equation}
collect logical prompt and output tokens, requests, uncached-prefill tokens,
time to first token, wall time, integrated and peak KV occupancy, peak
unified-memory pressure, and sampled GPU energy.  Here
$K_{\int}$, the KV time-area, integrates allocated KV bytes over wall time
(GiB$\cdot$s); unlike $K_{\max}$, it does not let one long failed request
define an arm.  With $\operatorname{LCB}_{.95}$
denoting the lower endpoint of a paired 95\% interval obtained by resampling
whole repositories, a policy is eligible only if
\begin{equation}
  \operatorname{LCB}_{.95}\!\left[Q(p)-Q(c(p))\right] > -\epsilon,
  \qquad \epsilon=0.05.
  \label{eq:guardrail}
\end{equation}
The control is \textsc{Grep} for the two localization treatments and
\textsc{Eager} for the \textsc{Mask-5} patch contrast.  The constraint is
applied after each experiment to decide which policies enter the cost
comparison; it does not act at runtime.  Failed tasks remain in denominators and
server resource totals, including requests without a serialized cell.  We also
report cost per successful endpoint so a cheap failure does not appear
efficient.

\section{Evaluation}
\label{sec:evaluation}

\paragraph{Questions.}
The RQs map to Section~\ref{sec:introduction}.  \textbf{RQ1 (C1/C3)} asks which
delivery and retention policies satisfy the quality gate and how they change
the agent trajectory.  \textbf{RQ2 (C2)} asks when the logical-token ordering
survives translation into prefill, KV occupancy, latency, memory, and energy.
A same-model prefix-cache intervention isolates prefix reuse.  \textbf{RQ3 (C1/C3)}
tests whether the token--work relationship recurs in an external agent
scaffold with official patch grading.

\paragraph{Localization Cohorts.}
Two disjoint cohorts share the CodeNib runner, retriever, policies, model,
and Recall@5 endpoint.  The synthesis cohort supports repeated device
measurement; the SWE-rebench-derived cohort powers the quality gate.

The resource cohort comes from the frozen \texttt{codenib-synthesis} test
split~\citep{codenib}.  It contains 25 pinned repository snapshots, five each for Python, Go,
Rust, TypeScript/JavaScript, and C/C++.  Category-balanced round robin selects
four of six issue-like query types per snapshot: behavioral, file hint, module
hint, symbol hint, reasoning, and traversal.  The resulting 100 tasks have
function/class answer blocks for Recall@5; three policy repetitions produce
900 cells.  Bootstrap draws retain each repository's tasks, repetitions, and
policy outcomes: the 300 cells per arm form 25 independent groups.
The dataset labels its queries LLM-curated and LLM-judged but omits both model
identifiers; we make no claim about human-authored issue distributions.

The powered quality cohort is derived from SWE-rebench
V2~\citep{swerebenchv2}, a public GitHub issue--patch collection.  Before model
calls, we select 450 Python issues from distinct repositories.  Eligible cases
have an applicable gold patch, at most 5,000 tracked Python files, and a
base-visible function/class block touched by the patch.  We use those blocks
as localization targets, separately from the SWE-bench Verified patch study.
Variance from the 100-task instrumented pilot gives a maximum requirement of
441 independent cases, rounded to 450.  One run per policy yields 1,350 cells and powers quality only;
physical costs use the 100-task cohort.  We do not pool the cohorts.

\paragraph{Localization Stack and Host.}
All localization studies, comprising the 450-instance quality cohort, 900-cell
resource matrix, and cache intervention, use the same Qwen3.6-35B-A3B-FP8 revision (35B total, 3B
active)~\citep{qwen36} with vLLM~\citep{vllm}.  We set temperature to zero, cap
context at 65,536 tokens, and admit one measured request at a time.  Prefix
caching is disabled in the resource matrix; the mechanism study toggles only that
flag.  A revision-pinned Qwen3-Embedding-0.6B server supplies every dense
top-10.  The host is a DGX Spark with an NVIDIA GB10 SoC and 128 GB coherent
unified memory~\citep{dgxspark}.  Appendix~\ref{app:protocol} records exact
revisions, serving versions, memory reservations, and telemetry metadata.

\paragraph{Cache Mechanism Study.}
RQ2 toggles one vLLM prefix-cache flag over the same 100 localization tasks.
Model, runtime, decoder, task order, and revision remain fixed, yielding 600
cells and 25 paired repositories per policy and regime.  Resets and salts
restrict reuse to each four-query block; an audit confirms that the serving
commands of the two regimes differ only in that flag.

\paragraph{External Patch Study and Stack Boundary.}
A repository-balanced seeded hash selects 100 SWE-bench Verified instances
across all 12 Python repositories~\citep{swebench,swebenchverified}, without inspecting gold
patches, tests, or outcomes.  Mini-SWE-agent~\citep{minisweagent} v2.4.6 uses its official prompt,
bash interface, and submission protocol.  A separately deployed SGLang
RadixAttention stack~\citep{sglang} uses DFlash speculative decoding with eight
draft tokens~\citep{dflash}; Appendix~\ref{app:protocol} gives both checkpoint
revisions.  We compare \textsc{Mini}, \textsc{Eager}, \textsc{Compact}, and
\textsc{Mask-5}; the three seeded arms receive identical top-10 candidates.
Each arm gets one trajectory, 32 calls, 1,800 seconds, temperature zero, and no
reference-patch access.  The stricter-than-upstream call cap makes resolved
rate a fixed-budget endpoint.

The external target and serving stack differ from localization, so we
compare policies only within each stack.  This study tests whether the pattern
recurs; it cannot separate the effects of target model, scaffold, and
runtime.

The revision-pinned official harness scores every attempt.  Empty patches and
failures remain in the denominator.  Generation and retrieval enter resource
totals; AMD64 grading does not.  Eight shards repeat each arm order twice, with
a Radix-cache reset per policy--shard block.

\paragraph{Controls.}
Localization arms share the base prompt, tools, 16-turn limit, and decoder.
A repository-level Latin square interleaves repetitions.  The patch study uses
the declared four-arm rotation.  Each measured block follows an unmeasured
warm-up and idle window; failed requests remain charged.

\paragraph{Measurements.}
Agent traces provide prompt tokens, completion, calls, and wall time.  Server
counters provide generation, prefill, TTFT, latency, and KV usage.  We report
\emph{successful server calls per attempted task}; ``per call'' always divides
by that server counter.  One-second KV sampling gives time-integrated
occupancy; SGLang counters also separate device hits from uncached prefill and
confirm DFlash activity.  We compute the GPU-energy proxy by trapezoidally
integrating sampled \texttt{nvidia-smi power.draw} over the run and
subtracting the pre-run mean idle power multiplied by the run duration.  It is
not whole-system energy.  Unified host memory comes from
\texttt{/proc/meminfo} and cgroups.  Prompt, wall time, and agent calls are per
cell.  Server counters, KV time-area, and power are aggregated per
policy--shard block in the primary and external patch studies, and per
policy--repository block in the same-model cache study.

\paragraph{Analysis.}
Full answer-block Recall@5 defines localization success.  A 10,000-draw paired
bootstrap averages repetitions and resamples 25 repository groups.  The
450-case cohort has one instance per repository, so its two resampling views
coincide; SWE-bench draws instead resample 12 source repositories.  Wall-time
intervals use 25 groups.  Block-level counters resample three disjoint shards
while retaining repetitions.  The instrumented study lacked prospective power
analysis.  We report build cost separately as $C_{\mathrm{build}}/n$.

\section{Results}
\label{sec:results}

The powered quality study attempted 1,350 cells on 450 instances.
Separately, the instrumented primary attempted 900 cells; its three runtime
failures are analyzed below.  Every attempted cell remains in its study denominator.

\begin{figure}[t]
  \centering
  \includegraphics[width=\linewidth]{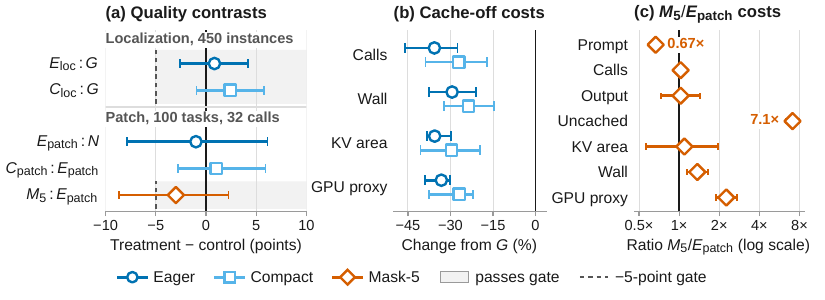}
  \caption{Quality gates and resource effects.  (a) Quality contrasts
  $\Delta Q_{p:c}=Q(p)-Q(c)$ for localization ($n=450$ instances) and patch
  generation ($n=100$, 32-call budget); only $E_{\mathrm{loc}}{:}G$,
  $C_{\mathrm{loc}}{:}G$, and $M_5{:}E_{\mathrm{patch}}$ face the gate.  (b)
  Cache-off effects relative to $G$; wall resamples 25 repositories and the
  other metrics resample three run-block shards.  (c) $M_5/E_{\mathrm{patch}}$
  resource ratios over eight paired shards.  $G/N$ receive no retrieved seed.
  Whiskers are paired-bootstrap 95\% intervals.}
  \label{fig:evidence}
\end{figure}

\subsection{RQ1: Quality-Eligible Policies and Shorter Trajectories}

\begin{table}[t]
  \centering
  \caption{Localization-only profile for
  $\mathcal{P}_{\mathrm{loc}}=\{G,E,C\}$.  Quality uses 450 instances;
  resources use the 100-task, three-repeat cache-off study.  ``Full'' means
  recall one.  Parenthesized changes are relative to $G$ (green: reduction);
  failures remain charged.
  Per-full-recall rows divide by the full-recall rate of the 100-task
  instrumented study.}
  \label{tab:profile}
  \small
  \setlength{\tabcolsep}{9pt}
  \renewcommand{\arraystretch}{1.02}
  \begin{tabular}{@{}lrrr@{}}
    \toprule
    Metric & \textsc{Grep} & \textsc{Eager} & \textsc{Compact} \\
           & \vsbase{control} & \vsbase{vs.\ $G$} & \vsbase{vs.\ $G$} \\
    \midrule
    \multicolumn{4}{@{}l}{\textit{Task quality}} \\
    Recall@5 & .470 & .478 & .494 \\
Full@5 & .307 & .318 & .329 \\
Format success & .904 & .976 & .987 \\
Runtime completion & 1.000 & 1.000 & 1.000
\tabularnewline
    \addlinespace[2pt]
    \multicolumn{4}{@{}l}{\textit{Resource per attempted task}} \\
    Prompt ktok               & 136.8 & 90.5 \saving{-33.9\%} & 98.2 \saving{-28.2\%} \\
    Successful server calls   & 13.02 & 8.38 \saving{-35.7\%} & 9.49 \saving{-27.1\%} \\
    Wall seconds              & 68.32 & 48.22 \saving{-29.4\%} & 52.20 \saving{-23.6\%} \\
    KV GiB$\cdot$s           & 25.88 & 16.67 \saving{-35.6\%} & 18.21 \saving{-29.6\%} \\
    GPU-proxy kJ              & 2.007 & 1.340 \saving{-33.2\%} & 1.466 \saving{-27.0\%} \\
    \addlinespace[2pt]
    \multicolumn{4}{@{}l}{\textit{Resource per full-recall localization}} \\
    Prompt ktok/full          & 204.2 & 133.7 \saving{-34.5\%} & 151.9 \saving{-25.6\%} \\
    GPU-proxy kJ/full         & 3.00  & 1.98 \saving{-34.0\%}  & 2.27 \saving{-24.3\%} \\
    \bottomrule
  \end{tabular}
\end{table}

Across 450 localization cases derived from SWE-rebench V2, one per repository,
the frozen retriever reaches
\PoweredRetrieverRecallFive{} Recall@5 and \PoweredRetrieverRecallTen{}
Recall@10.  Agent Recall@5 is \PoweredGrepRecall, \PoweredEagerRecall, and
\PoweredCompactRecall{} for \textsc{Grep}, \textsc{Eager}, and
\textsc{Compact} (Table~\ref{tab:profile}).  Nonconforming outputs remain in
the denominator and score zero.

For \textsc{Eager}, the paired difference is $\PoweredEagerDiffPoints$ points;
its lower 95\% endpoint is $\PoweredEagerCILowPoints$ and clears the
$-5$-point gate.  \textsc{Compact} differs by
$\PoweredCompactDiffPoints$ points, with lower endpoint
$\PoweredCompactCILowPoints$.  It \PoweredCompactGate{} the same criterion
(Figure~\ref{fig:evidence}a).

In the instrumented primary, \textsc{Eager} saves 46.3k prompt tokens per task,
or 33.9\%; the paired interval ranges from 28.8k to 65.4k saved.
\textsc{Compact} saves 28.2\% across all attempts.  Its 99 runtime-complete
pairs save 39.8k tokens on average, with endpoints at 21.6k and 58.0k saved.
The useful comparison for history reduction is \textsc{Compact} against its
retrieval-matched \textsc{Eager} control.  \textsc{Compact} uses 8.5\% more
prompt tokens, 13.3\% more calls, 8.2\% more wall time, and 9.4\% more energy.
Its requests are not longer: dividing prompt volume by successful calls gives
at most 10.3k tokens per call for \textsc{Compact} and 10.8k for
\textsc{Eager}.  Prefix caching is off in this matrix, so the extra calls
account for the gap.  The anchor drops the remaining candidates, and the agent
may search again for evidence that \textsc{Eager} still shows; traces do not
label call intent, so this attribution remains open.

\subsection{RQ2: Physical Savings Follow Fewer Calls}
For all four primary cost metrics, the intervals for both policies exclude
zero (Figure~\ref{fig:evidence}b).
Normalizing energy by full-recall localizations yields 3.00, 1.98, and 2.27 kJ
for \textsc{Grep}, \textsc{Eager}, and \textsc{Compact}.  The quality-adjusted
ordering therefore matches the per-attempt ordering (Table~\ref{tab:profile}).

Aggregate energy divided by wall time stays within 27.7--29.3\,W in the primary
and 25.2--26.0\,W in the SGLang stack, so energy mainly tracks duration; calls,
uncached prefill, and KV time-area show why costs differ.  Mean TTFT varies by
less than 2\%, and latency per successful call rises from 5.218 to 5.721
seconds (\textsc{Eager}) and 5.468 (\textsc{Compact}), so the task-time saving
comes from fewer calls.  Peak-KV and system-memory changes are too noisy to
support a capacity claim (Appendix~\ref{app:protocol}).

\paragraph{Failures.}
One MicroPython \textsc{Compact} task fails in all repetitions because 61,441
prompt plus 4,096 output tokens exceed the server's context limit by one; KV
memory never runs out.  The headline retains its 1.204M failed-request tokens:
98.2k per attempt and 28.2\% savings.  Removing that charge gives 95.1k over
297 completed cells and 30.5\% savings.

\paragraph{Build Cost and Cache Isolation.}
Compiling the 25 repository indexes takes 767.7 seconds and 36.1 kJ, producing
147,387 chunks in 731 MB.  At ten queries per snapshot, amortization adds 3.07
seconds and 0.144 kJ per query; at 100 queries its energy share is below 1.1\%.

In the 600-cell cache intervention, turning caching on changes \textsc{Eager}'s logical prompt by
$\CacheEffectEagerPromptChange$\% (95\% CI
[$\CacheEffectEagerPromptCILow$, $\CacheEffectEagerPromptCIHigh$]) but its
uncached prefill by $\CacheEffectEagerUncachedChange$\%
([$\CacheEffectEagerUncachedCILow$, $\CacheEffectEagerUncachedCIHigh$]).
The planned $C_{\mathrm{loc}}-E_{\mathrm{loc}}$ cache interaction is
$\CacheCompactVsEagerUncachedKInteraction$k uncached tokens per task; its
95\% interval runs from $\CacheCompactVsEagerUncachedKInteractionCILow$k to
$\CacheCompactVsEagerUncachedKInteractionCIHigh$k.  The wall interaction is
$\CacheCompactVsEagerWallSecondsInteraction$ seconds, ranging from
$\CacheCompactVsEagerWallSecondsInteractionCILow$ to
$\CacheCompactVsEagerWallSecondsInteractionCIHigh$.  Both interaction
intervals include zero.  The same-model cache-on \textsc{Compact} contrast
against \textsc{Grep} is $\CacheOnCompactUncachedChange$\% uncached prefill,
so the $+5.2\%$ increase in the exploratory Qwen3.8/SGLang stack
(Table~\ref{tab:secondary-absolute}) does not reproduce.

\subsection{RQ3: Token Savings Reverse under External Retention}

\begin{table}[t]
  \centering
  \caption{Patch-generation-only profile for $\mathcal{P}_{\mathrm{patch}}$:
  mini-SWE-agent on 100 SWE-bench Verified instances under a 32-call budget.
  Resources are generation-only means per attempt; parenthesized changes use
  the baseline under each arm (green: reduction; red: increase).}
  \label{tab:mini-e2e}
  \small
  \setlength{\tabcolsep}{7pt}
  \renewcommand{\arraystretch}{1.02}
  \begin{tabular}{@{}lrrrr@{}}
    \toprule
    Metric & \textsc{Mini} & \textsc{Eager} & \textsc{Compact} & \textsc{Mask-5} \\
           & \vsbase{control} & \vsbase{vs.\ $N$} & \vsbase{vs.\ $E$}
           & \vsbase{vs.\ $E$} \\
    \midrule
    \multicolumn{5}{@{}l}{\textit{Task outcome (of 100 attempts)}} \\
    Resolved & 30 & 29 & 30 & 26 \\
Submitted / call-capped & 45 / 55 & 44 / 56 & 45 / 55 & 40 / 60 \\
\addlinespace[2pt]
\multicolumn{5}{@{}l}{\textit{Generation resource per attempt}} \\
Prompt ktok & 220.7 & 225.6~\increase{+2.2\%} & 210.8~\saving{-6.5\%} & 150.5~\saving{-33.3\%} \\
Model calls & 26.13 & 25.66~\saving{-1.8\%} & 25.84~\increase{+0.7\%} & 26.33~\increase{+2.6\%} \\
Uncached ktok & 9.8 & 10.0~\increase{+1.4\%} & 12.2~\increase{+22.0\%} & 70.8~\increase{+608.5\%} \\
KV frac.$\cdot$s & 2.94 & 3.25~\increase{+10.6\%} & 2.69~\saving{-17.2\%} & 3.55~\increase{+9.4\%} \\
Wall seconds & 120.2 & 122.7~\increase{+2.0\%} & 116.6~\saving{-5.0\%} & 167.9~\increase{+36.9\%} \\
GPU-proxy kJ & 1.833 & 1.956~\increase{+6.7\%} & 1.917~\saving{-2.0\%} & 4.412~\increase{+125.6\%}
\tabularnewline
    \bottomrule
  \end{tabular}
\end{table}

Audits confirm that all \MiniValidatedNoSeedCells{} no-seed cells received no
seed and all \MiniValidatedPreloadCells{} preload cells match the frozen
candidate and query hashes.  Thus \textsc{Eager}--\textsc{Mask-5} changes observation
visibility with retrieval fixed.  Table~\ref{tab:mini-e2e} accounts for every
trajectory as a submission or a 32-call cap, and no infrastructure failure
enters the result.
Across \MiniRepositoryClusters{} source repositories, \textsc{Mask-5}
resolves \MiniMaskResolved/\MiniMaskAttempts{} attempts versus
\MiniEagerResolved/\MiniEagerAttempts{} for \textsc{Eager}.  Its paired
difference is $\MiniMaskDiffPoints$ points.  The lower 95\% endpoint,
$\MiniMaskCILowPoints$, crosses the $-5$-point
gate, so it \MiniMaskQualityGate{} the non-inferiority rule
(Figure~\ref{fig:evidence}a).  Masking activates at median call
\MiniMaskFirstCallMedian{}.  By Equation~\eqref{eq:cost-decomposition},
\textsc{Mask-5} removes 75.1k logical prompt tokens per attempt but
loses 135.9k device-hit tokens, adding 60.8k uncached tokens.  Cached prefixes
serve about 96\% of \textsc{Eager}'s prompt but 53\% of \textsc{Mask-5}'s.
Aggregate prompt
volume changes by $\MiniMaskPromptChange$\% and uncached prefill by
$\MiniMaskUncachedChange$\%; both directions hold across all eight paired
shards.  Calls change by $\MiniMaskCallsChange$\%, wall time by
$\MiniMaskWallChange$\%, and the GPU-energy proxy by $\MiniMaskEnergyChange$\%.
The prefill inversion is consistent with reduced prefix-cache reuse after
repeated history edits.  Trajectory changes may also contribute to the
observed resource differences.

\section{Implications for Resource-Aware Context Management}
\label{sec:implications}

\paragraph{Separate Evidence Delivery from Retention (C1).}
Eager uses the fewest resources among eligible localization arms.  The anchor
remains eligible but adds no saving, because shorter requests do not offset
its additional calls.

\paragraph{Measure Prefix Reuse (C2).}
Equation~\eqref{eq:cost-decomposition} also locates the break-even point for
history deletion: a policy lowers uncached prefill only if it removes more
logical tokens than reused ones.  \textsc{Mask-5} removes 75.1k and loses
135.9k.  As a first-order estimate, if a runtime served a fraction $\rho$ of
the hits observed here in both arms, masking would change uncached prefill by
$-75.1\mathrm{k}+135.9\mathrm{k}\cdot\rho$ tokens per attempt, which turns
positive above $\rho\approx0.55$.  At $\rho=0$, without a prefix cache,
masking would save prefill; batching or cache pressure that evicts prefixes
lowers $\rho$ and with it the masking penalty.  Reports should therefore expose request
count, prefix reuse, and uncached prefill alongside prompt volume.

\paragraph{Constrain Quality Before Comparing Resources (C3).}
Apply Equation~\eqref{eq:guardrail} before interpreting resource differences.
$E_{\mathrm{loc}}$ and $C_{\mathrm{loc}}$ meet it against $G$, while $M_5$
\MiniMaskQualityGate{} it against $E_{\mathrm{patch}}$; token count alone
would favor the policy that the gate excludes.  Other patch contrasts remain
descriptive.

\section{Related Work}
\label{sec:related}

\paragraph{Repository Localization and Agent Context.}
BugLocator first combined report--code similarity with development history
\citep{buglocator}; BLUiR then indexed code structure explicitly~\citep{bluir},
and learning-to-rank incorporated API, history, and method features
\citep{learningtorankbug}.  RepoCoder moved repository retrieval into an
iterative generation loop~\citep{repocoder}, while RepoBench measured broader
repository context~\citep{repobench}.  SWE-bench made localization an upstream
step in executable issue resolution~\citep{swebench}.  AutoCodeRover exposed
AST-aware search~\citep{autocoderover}; Agentless used hierarchical
file-to-function localization~\citep{agentless}.  LocAgent, CoSIL, RepoGraph,
and CodexGraph navigate repository code graphs
\citep{locagent,cosil,repograph,codexgraph}; OrcaLoca prioritizes search
actions and prunes context by code distance~\citep{orcaloca}.  We adopt the resulting function/class endpoint,
freeze retrieval, and study the resources used to deliver and retain evidence.

Static prompt compression precedes long-horizon agent memory
\citep{llmlingua,longllmlingua}.  Later policies externalize memory, mask or
summarize observations, prune lines, and maintain structured state
\citep{memgpt,complexity-trap,agentdiet,context-as-tool,swe-pruner,codemem}.
Repository compression also varies by task~\citep{repo-context-compression}.

\paragraph{Runtime Context and KV Reuse.}
PagedAttention allocates KV blocks; RadixAttention, Prompt Cache, RAGCache, and
CacheBlend reuse state~\citep{vllm,sglang,promptcache,ragcache,cacheblend}.
CacheGen transfers compressed state, while H$_2$O, StreamingLLM, and SnapKV
retain selected tokens~\citep{cachegen,h2o,streamingllm,snapkv}.  ContextPilot
aligns reusable context; Stream2LLM models updates that invalidate cache
\citep{contextpilot,stream2llm}.  KV compression can also lengthen outputs and
latency~\citep{rethinkkv}.  We measure how context policies interact with
this reuse.

\paragraph{Resource-Aware Evaluation.}
Green AI and HELM motivate cost-aware, multi-metric evaluation; Zeus measures
GPU energy directly~\citep{greenai,helm,zeus}.  Our protocol adds a prespecified
quality gate, uncached prefill, and KV time-area to these metrics and charges
failures and index construction.

\section{Limitations and Conclusion}
\label{sec:limitations}

Results cover two stacks on one DGX Spark serving one request at a time;
batching, cache pressure, and cloud GPUs remain untested.  The Spark exposes
no application-clock or power-limit control (\texttt{power.limit=N/A}), so
clocks were not pinned; Latin-square interleaving spreads drift across
policies, and average power stays nearly constant across arms.  Server, KV,
and energy intervals use only three shards.  Quality and detailed cost come
from different localization cohorts (450 and 100 tasks), so no single issue set
carries both measurements.  The patch study covers
100 Python tasks from 12 repositories, with physical intervals over eight
shards.  Powered
localization has one trajectory per policy per repository.  The 32-call patch
cap measures fixed-budget performance, not unrestricted SWE-bench ability.
We test four fixed policies; \textsc{Mask-5} fixes the window at five
observations, following the SWE-agent reference.  Window sweeps, LLM
summarization, learned pruning, and external memory remain unmeasured, although
Equation~\eqref{eq:cost-decomposition} applies to each.  Mini-SWE-agent lacks
observation tags; SGLang reports only full-attention KV occupancy.  Our GPU
proxy omits CPU, host-memory, and whole-system energy.  On-device
execution retains source privacy, but cheaper misuse and incorrect patches
remain risks requiring sandboxing and human review.

\paragraph{Conclusion.}
A context policy changes both what the agent does and how much of its prompt
the runtime can reuse.  Early evidence saved work by avoiding calls, while
repeated observation masking turned a 33.3\% prompt saving into a 608.5\% rise
in uncached prefill.  Resource comparisons should measure executed work and
admit only policies that pass a quality gate.

\clearpage
\label{raaai:references-start}
\bibliography{references}
\bibliographystyle{unsrtnat}

\appendix
\section{Frozen Protocol and Artifact Notes}
\label{app:protocol}

The primary configuration pins the implementation commit, two dataset
revisions, task-selection rule, model endpoint, decoding parameters, and policy
definitions.  Before each run, the artifact captures the host, driver, kernel,
Python packages, serving command, model identity, current processes, and
repository status.  Raw agent outputs and telemetry are immutable; derived
tables are regenerated from them.

Table~\ref{tab:execution-metadata} records the execution metadata captured with
the frozen runs.  Application clocks and a configurable device power limit were
not available in the capture: \texttt{nvidia-smi} reported
\texttt{power.limit=N/A}.  The study therefore makes no clock-stability claim.

\begin{table}[!htbp]
  \centering
  \caption{Frozen execution metadata.  Readable versions and abbreviated
  revisions are shown here; the artifact manifest records full revisions and
  integrity checks.}
  \label{tab:execution-metadata}
  \footnotesize
  \setlength{\tabcolsep}{5pt}
  \renewcommand{\arraystretch}{1.12}
  \begin{tabular}{@{}>{\raggedright\arraybackslash}p{0.17\linewidth}
                  >{\raggedright\arraybackslash}p{0.79\linewidth}@{}}
    \toprule
    Field & Captured value \\
    \midrule
    Host & NVIDIA GB10, 128 GB unified memory, \texttt{aarch64} \\
    OS / driver & Ubuntu 24.04.4 LTS; kernel \texttt{6.17.0-1026-nvidia}; NVIDIA driver 580.159.03; CUDA 13.0 \\
    Primary target & Qwen3.6-35B-A3B-FP8, revision \texttt{95a723d08a94} \\
    Primary runtime & vLLM \texttt{0.19.2rc1.dev134+gfe9c3d6c5} \\
    Cache intervention & Same Qwen3.6/vLLM stack; treatment adds only \texttt{-{}-enable-prefix-caching}, with a reset before each policy--repository block \\
    External target & Qwen3.8-27B-FP8~\citep{qwen38}, revision \texttt{017b9c7af6b5} \\
    External runtime & SGLang \texttt{0.0.0.dev1+gc0b6474b4}; image \texttt{codenib/sglang-dflash2:c14312a6-cu130-arm64} \\
    Draft model & Qwen3.8-27B-DFlash2, revision \texttt{dedf8df68adf}; eight draft tokens \\
    Retriever & Qwen3-Embedding-0.6B, revision \texttt{97b0c614be4d}; vLLM runtime above \\
    Agent code & CodeNib revision \texttt{88e88b38b978} \\
    External scaffold & mini-SWE-agent v2.4.6 (\texttt{a83fcae82d2a}); official SWE-bench harness (\texttt{7a21e0577295}) \\
    Observation masking & SWE-agent \texttt{LastNObservations} reference (\texttt{3ea751c087f3}); last $n=5$, polling${}=1$ \\
    Power samples & \texttt{nvidia-smi power.draw}, nominal 1 s interval; 30 s primary/E2E and 5 s cache-intervention pre-run idle windows \\
    Idle baseline & Primary: mean 12.278 W, range 10.932--12.735 W over 27 blocks; exploratory cached stack: mean 12.085 W, range 11.529--12.354 W over nine blocks.  Cache-intervention and external-E2E energy use the corresponding per-block pre-run mean retained in each telemetry artifact. \\
    Resident services & Xorg and GNOME Shell; target and embedding servers.  No other client accessed either endpoint during formal blocks. \\
    \bottomrule
  \end{tabular}
\end{table}

The run sequence is: (1) rebuild or verify repository artifacts; (2) validate
the three arms on two tasks and audit compaction traces; (3) perform unmeasured
model warm-up; (4) execute three repository shards in rotated policy order;
(5) repeat the rotation twice; and (6) bootstrap paired task differences by
repository.  A shard is invalidated if another client uses the model endpoint,
the model process restarts, telemetry has a gap longer than two sampling
intervals, or memory pressure triggers swapping.  Invalidated shards remain
disclosed before reruns; no shard is deleted.

The post-run audit resolves and verifies all 897 persisted traces.  It retains
a failed status for the other three cells because the exception serializer
omits their preload-candidate hash along with the final trace record.  Those
are the same disclosed context-limit failures; their task and candidate set
remain fixed by the selection manifest.  We preserve this secondary audit
failure without rewriting raw artifacts.

For energy, samples $(t_i,P_i)$ are integrated as
\begin{equation}
 E_{\mathrm{raw}}=\sum_i \frac{P_i+P_{i+1}}{2}(t_{i+1}-t_i), \qquad
 E_{\mathrm{proxy}}=E_{\mathrm{raw}}-\bar{P}_{\mathrm{idle}}\Delta t.
\end{equation}
Negative idle-subtracted values remain measurement noise in the statistical
input; no sample is silently clipped.  Whole-system energy requires an
external power meter and is outside the primary protocol.

Formal primary and exploratory matrices consume 14.11 and 4.23 measured
device-hours, respectively.  Compilation adds 0.213 hours.
The same-model cache intervention adds \CacheInstrumentedHours{} device-hours
across \CacheInstrumentedBlocks{} telemetry blocks, while external
mini-SWE-agent generation adds \MiniInstrumentedHours{} device-hours across
\MiniInstrumentedBlocks{} blocks.  Their idle-subtracted GPU-energy proxies
are \CacheInstrumentedEnergyKWh{} and \MiniInstrumentedEnergyKWh{} kWh,
respectively.  These two added studies therefore account for
\AddedInstrumentedHours{} device-hours and \AddedInstrumentedEnergyKWh{} kWh;
official SWE-bench grading is excluded.
The accounting excludes model-server residency and other uninstrumented
development.
A normal DGX Spark graphics session (Xorg and the desktop compositor) remained
resident throughout; the idle window captures that background.  Only the agent
target and embedding containers issued compute requests during formal shards,
and neither endpoint received an external client request.

For the exploratory cached-stack probe, all nine Radix-cache resets return HTTP
200, all 300 cells complete,
the maximum one-second telemetry gap is 1.16 seconds, and no request queues or
swap growth occur.  Each arm records nonzero DFlash verification calls.  The
four-task smoke test is not used as paper evidence.

The external E2E protocol uses the same cached target, draft model, retrieval
service, and 65,536-token server limit.  It runs \textsc{Mini}, \textsc{Eager}, \textsc{Compact}, and
\textsc{Mask-5} over eight near-equal shards.  Every arm occupies each order position
twice, and the Radix cache is reset before every policy--shard block.  The
generation limit is 32 model calls.  Official AMD64 grading is excluded from
generation-resource totals, while empty patches and test failures remain in
the 100-instance denominator.

One pre-execution sandbox check rejected an empty, uninitialized Astropy
gitlink.  We quarantined and invalidated its entire
\MiniRecoveryInvalidatedCells-cell \textsc{Eager} block, including
\MiniRecoveryDiscardedRequests{} requests and
\MiniRecoveryDiscardedMinutes{} measured minutes, then repeated the block
after a fresh cache reset.  The recovery permits only revision-bound gitlinks
whose worktree directories are empty and uninitialized; networking remains
disabled.  Artifact-preserved copies of the failed cell, discarded telemetry,
run manifest, predictions, and original pipeline status remain outside every
aggregate.

\begin{table}[!htbp]
  \centering
  \caption{Exploratory external SGLang stack (100 tasks, one repetition).  Prompt and
  uncached-prefill values are SGLang counters per attempted task; KV is
  integrated normalized full-attention-pool occupancy; energy is the same
  idle-subtracted GPU proxy as in the main paper.}
  \label{tab:secondary-absolute}
  \footnotesize
  \setlength{\tabcolsep}{6pt}
  \begin{tabular}{@{}lrrrrrrr@{}}
    \toprule
    Policy & Recall@5 & Prompt & Uncached & Req. & Wall & KV \%$\cdot$s & kJ \\
    \midrule
    \textsc{Grep}    & .772 & 98.779k & 18.245k & 10.34 & 56.76 & 165.8 & 1.474 \\
    \textsc{Eager}   & .800 & 68.456k & 13.749k &  7.15 & 42.71 & 118.9 & 1.080 \\
    \textsc{Compact} & .802 & 84.226k & 19.192k &  8.87 & 52.16 & 133.8 & 1.362 \\
    \bottomrule
  \end{tabular}
\end{table}

\begin{table}[!htbp]
  \centering
  \caption{Per-successful-server-call decomposition.  Final is the fraction of captured trace
  calls that produce the final answer; output and TTFT use aggregate server
  counters.  E2E/call is mean server end-to-end latency per successful model
  request.  In this table, uncached prefill is reported only for the
  exploratory SGLang rows.  These
  ratios are descriptive, not paired estimates.}
  \label{tab:per-call}
  \footnotesize
  \setlength{\tabcolsep}{6pt}
  \begin{tabular}{@{}llrrrrr@{}}
    \toprule
    Regime & Policy & Final \% & Output/call & Uncached/call & TTFT s & E2E/call s \\
    \midrule
    \multirow{3}{*}{Primary}
      & \textsc{Grep}    &  7.9 & 160.4 & --     & 2.288 & 5.218 \\
      & \textsc{Eager}   & 12.0 & 190.2 & --     & 2.312 & 5.721 \\
      & \textsc{Compact} & 10.7 & 182.3 & --     & 2.244 & 5.468 \\
    \midrule
    \multirow{3}{*}{\shortstack[l]{Exploratory\\(SGLang)}}
      & \textsc{Grep}    &  9.7 & 126.2 & 1.765k & 4.049 & 5.410 \\
      & \textsc{Eager}   & 14.0 & 137.7 & 1.923k & 4.161 & 5.840 \\
      & \textsc{Compact} & 11.3 & 126.5 & 2.164k & 4.288 & 5.692 \\
    \bottomrule
  \end{tabular}
\end{table}

Table~\ref{tab:per-call} decomposes output, time to first token, and latency
per successful server call, and Table~\ref{tab:repository-pins} lists the 25
pinned repository snapshots.  Percentages in the main text and figures use
unrounded counters.  Displayed
table entries are rounded only after division.

The artifact also records compilation time and energy.  For an assumed $n$
queries against one snapshot, amortized cost is reported as
$C_{\mathrm{query}}+C_{\mathrm{build}}/n$ for $n\in\{1,10,100,1000\}$.

\begin{table}[!htbp]
  \centering
  \caption{Repository snapshots used by the frozen localization matrix, five per
  language group (Py: Python; TS/JS: TypeScript/JavaScript).  Revisions are
  abbreviated; the artifact manifest stores full commits.}
  \label{tab:repository-pins}
  \scriptsize
  \setlength{\tabcolsep}{4pt}
  \renewcommand{\arraystretch}{1.08}
  \begin{tabular}{@{}llll@{\hspace{14pt}}lll@{}}
    \toprule
    Repository & Lang. & Commit && Repository & Lang. & Commit \\
    \cmidrule(r){1-3}\cmidrule(l){5-7}
    \texttt{astropy/astropy} & Py & \texttt{0df94ff70979} && \texttt{tokio-rs/tokio} & Rust & \texttt{ab53bf0c4727} \\
    \texttt{matplotlib/matplotlib} & Py & \texttt{6786f437df54} && \texttt{uutils/coreutils} & Rust & \texttt{a40884758329} \\
    \texttt{pydata/xarray} & Py & \texttt{45c0a114e2b7} && \texttt{axios/axios} & TS/JS & \texttt{c6cce43cd944} \\
    \texttt{scikit-learn/scikit-learn} & Py & \texttt{97523985b39e} && \texttt{babel/babel} & TS/JS & \texttt{35116224d21e} \\
    \texttt{sympy/sympy} & Py & \texttt{d7c304511569} && \texttt{facebook/docusaurus} & TS/JS & \texttt{5e9e1d051b22} \\
    \texttt{caddyserver/caddy} & Go & \texttt{4512be49a9fa} && \texttt{preactjs/preact} & TS/JS & \texttt{c77e6284c199} \\
    \texttt{gin-gonic/gin} & Go & \texttt{9f598a31aafb} && \texttt{vuejs/core} & TS/JS & \texttt{67d6596d40b1} \\
    \texttt{gohugoio/hugo} & Go & \texttt{9c4e14eb4f89} && \texttt{fmtlib/fmt} & C/C++ & \texttt{840ec8569ddb} \\
    \texttt{hashicorp/terraform} & Go & \texttt{047c1af0b4a9} && \texttt{jqlang/jq} & C/C++ & \texttt{98f709d0c1fd} \\
    \texttt{prometheus/prometheus} & Go & \texttt{9f57f14d6c5e} && \texttt{micropython/micropython} & C/C++ & \texttt{b0ba151102a6} \\
    \texttt{astral-sh/ruff} & Rust & \texttt{b2a0d68d70ee} && \texttt{redis/redis} & C/C++ & \texttt{94b9072e44af} \\
    \texttt{nushell/nushell} & Rust & \texttt{fb34a4fc6c9c} && \texttt{valkey-io/valkey} & C/C++ & \texttt{4de8cf3567b4} \\
    \texttt{sharkdp/bat} & Rust & \texttt{2be3a14a7e2c} && & & \\
    \bottomrule
  \end{tabular}
\end{table}

\paragraph{Assets and Licenses.}
Apache-2.0 covers the agent code, the CodeNib synthesis dataset, and the
primary Qwen checkpoint; the pinned CodeNib base dataset is MIT-licensed.  The
Qwen3.8 target and DFlash draft checkpoints also use Apache-2.0, and the
SWE-rebench V2 dataset is CC-BY-4.0.  Each of the
25 evaluated repositories retains its individual upstream license and pinned
commit.  To avoid relicensing source-derived index contents, the artifact can
distribute manifests and rebuild scripts without bundling the compiled vectors.
Code and artifacts are released at \url{https://github.com/sysevol-ai/CodeNib}.

\end{document}